\documentclass[12pt]{iopart}

\usepackage{iopams}
\expandafter\let\csname equation*\endcsname\relax
\expandafter\let\csname endequation*\endcsname\relax
\usepackage{amsmath}  
\usepackage{amsfonts} 
\usepackage{siunitx} 
\usepackage{verbatim}
\usepackage{color}
\usepackage[english]{babel}
\usepackage{blindtext}
\usepackage[pdftex]{graphicx}
\usepackage{epstopdf} 
\usepackage{hyperref}
\usepackage{cite}
\usepackage[version=4]{mhchem} 
\newcommand{\YBCO}{\ce{YBa_2Cu_3O_{6+y}}}
\newcommand{\YBCOo}{\ce{YBa_2Cu_3O_{6.95}}}

\begin{document}

\title[The Superconducting Transition and Mixed State of \YBCOo]{The Superconducting Transition and Mixed State of \YBCOo: An Undergraduate Experiment} 

\author{Zhongda Huang, Yihang Tong and Jake S Bobowski} 
\address{Department of Physics, University of British Columbia, Kelowna, British Columbia, Canada V1V 1V7} 
\ead{jake.bobowski@ubc.ca}


\begin{abstract}
We describe a simple AC susceptometer built in-house that can be used to make high-resolution measurements of the magnetic susceptibility of high-temperature superconductors in an undergraduate physics lab.  Our system, cooled using liquid nitrogen, can reach a base temperature of \SI{77}{\kelvin}.  Our apparatus does not require gas handling systems or PID temperature controllers.  Instead, it makes use of a thermal circuit that is designed to allow the sample to cool on a time scale that is suitable for an undergraduate lab.  Furthermore, the temperature drift rate at the superconducting transition temperature $T_\mathrm{c}$ is low enough to allow for precise measurements of the complex magnetic susceptibility through $T_\mathrm{c}$, even for single-crystal samples with exceedingly sharp superconducting transitions.  Using an electromagnet, we were able to apply static magnetic fields up to \SI{63}{\milli\tesla} at the sample site.  By measuring the change in susceptibility as a function of the strength of an applied of static magnetic field, we were able to estimate the lower critical field $H_\mathrm{c1}$ of a single-crystal sample of optimally-doped \YBCOo{} at \SI{77}{\kelvin}.  We also investigated the mixed state of a sintered polycrystalline sample of \YBCO{}. 

~



\end{abstract}


\section{Introduction}\label{sec:intro}
AC susceptometry is an experimental technique widely used in research~\cite{Bidinosti:1999, Bidinosti:2000, Ahrentrop:2010, Chen:2004, Yonezawa:2015} and in undergraduate laboratories~\cite{Jaeger:1998, Perez:2018, Roy:2020}.  AC susceptometers are sensitive, but inexpensive, instruments that can be built in-house~\cite{Nikolo:1995, Gomory:1997}.  Furthermore, they can be used to investigate a variety of materials exhibiting a wide range of physical phenomena.  Examples of the systems studied by AC susceptometry include: spin ices~\cite{Snyder:2004}, spin glasses~\cite{Singh:2008}, superparamagnets~\cite{Bajpai:2000}, heavy fermions~\cite{Steppke:2013}, superconductors~\cite{Bidinosti:1999, Dhingra:1993, Sarmago:2004, Hicks:2014, Grinenko:2021} and systems exhibiting critical fluctuations at magnetic phase transitions~\cite{Berndt:1995}.

We describe a simple apparatus for the undergraduate laboratory designed to make sensitive measurements of the complex AC susceptibility of high-temperature superconductors at temperatures of \SI{77}{\kelvin} and above.  Thorough reviews of AC susceptibility measurements applied to high-temperature superconductors already exist in the literature~\cite{Nikolo:1995, Gomory:1997}.  This paper, therefore, focuses on the aspects of our apparatus and measurements that are unique.  Many undergraduate laboratories using AC susceptometers to investigate superconductivity focus on the measurement technique and an observation of the superconducting transition.  Our goal was to develop additional laboratory exercises that allow students to investigate the behaviour of the superconducting state.  Some authors have described using the temperature dependence of the measured susceptibility to determine the average grain size in sintered ceramic superconductors~\cite{Nikolo:1995, Gomory:1997, Jaeger:1998}. However, this analysis often requires measurements below \SI{77}{\kelvin}; a temperature range that can only be accessed using liquid $^4$He or a closed-cycle cryocooler~\cite{Radebaugh:2009} and is out of reach for many undergraduate labs.

One unique aspect of our apparatus is the use of a thermal circuit to cool the superconducting sample from room temperature to \SI{77}{\kelvin}.  The thermal circuit is designed to cool the sample to a stable base temperature within about \SI{2}{hours}.  Furthermore, at about \SI{1}{\kelvin/\minute}, the temperature drift rate through the superconducting transition temperature ($T_\mathrm{c}\approx\SI{93}{\kelvin}$ for \YBCOo{}) is low.  This allows for a detailed measurement of the temperature dependence of the AC susceptibility $\chi^\prime+\mathrm{i}\chi^{\prime\prime}$ through the superconducting transition.  Because the sample temperature is determined by the physical properties of the sample stage, our apparatus does not require a gas handling system, PID temperature controller or sample heater.  As will be discussed, we access the full temperature range from room temperature to \SI{77}{\kelvin} using only a liquid nitrogen bath, radiation shield, and vacuum chamber.

Because our apparatus is compact, the vacuum chamber can be inserted into the bore of an electromagnet enabling a measurement of the complex susceptibility as a function of static magnetic field strength at a fixed temperature.  In our experiments, we use an electromagnet, also submerged in the liquid nitrogen bath, to apply static fields up to \SI{63}{\milli\tesla} at the sample site.  This arrangement allows for an investigation of the mixed state of type-II superconductors in which there is a mixture of superconducting and normal state domains in the sample.  Furthermore, when working with high-quality single-crystals of \YBCOo{}, the onset field of the mixed state can be identified which allows for a reasonable an estimate of the lower critical field $H_\mathrm{c1}$.

\section{Apparatus design}
\begin{figure}[t]
\centering{
\includegraphics[keepaspectratio, width=0.9\columnwidth]{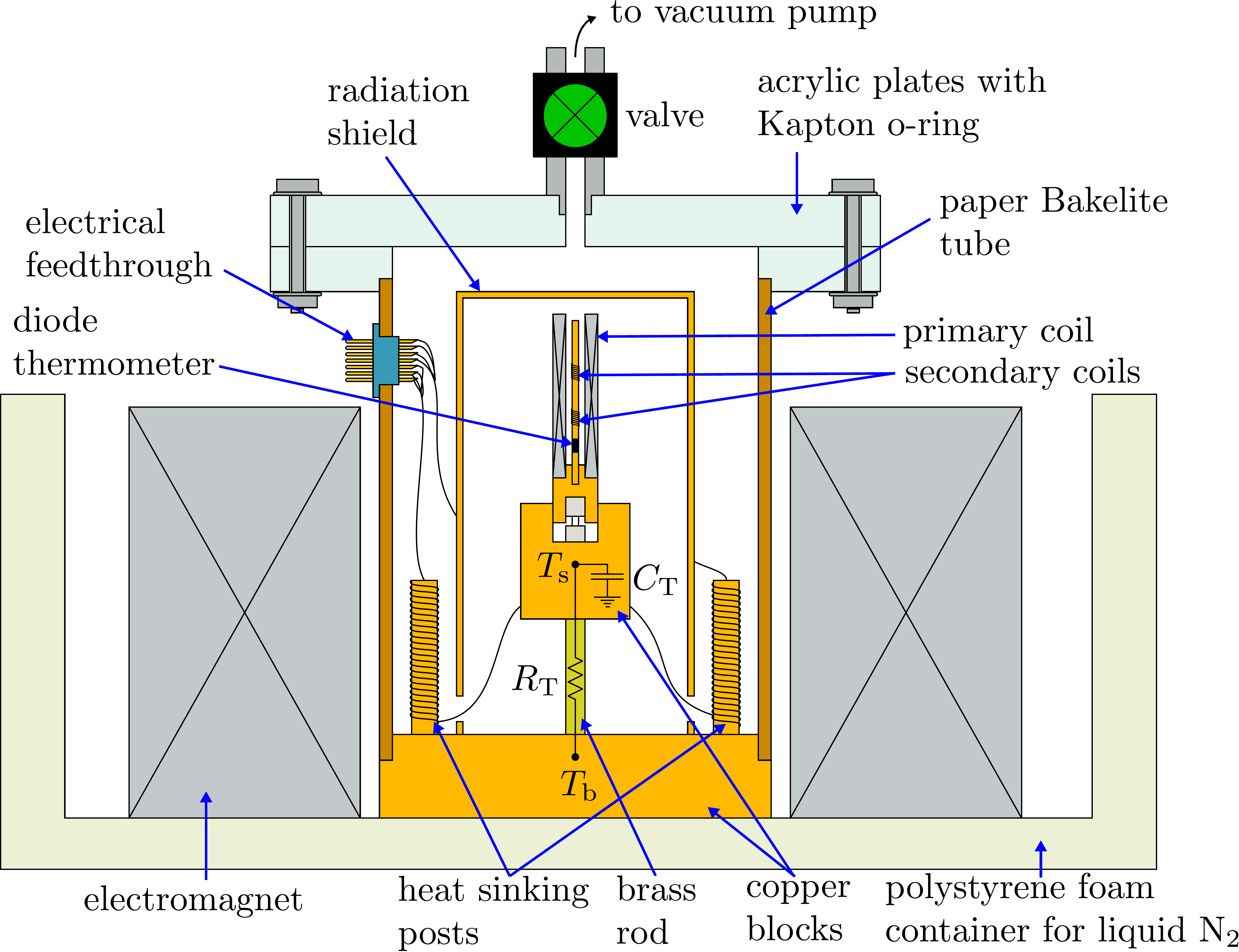}}
\caption{\label{fig:apparatus}Schematic diagram of the apparatus (not to scale).  The AC susceptometer is mounted on a thermal stage which sets the sample temperature $T_\mathrm{s}$.  The cooling time constant is set by the thermal capacitance $C_\mathrm{T}$ and resistance $R_\mathrm{T}$.  The susceptometer coils and thermal stage are surrounded by a copper radiation shield and contained inside a vacuum chamber.  Before connecting to the thermal stage, electrical leads are wrapped around heat sinking posts that are anchored to a copper plate that is in direct contact with the liquid nitrogen bath.  The vacuum chamber fits within the bore of an electromagnet that is also submerged in the liquid nitrogen bath.}
\end{figure}

\begin{figure}[t]
\centering{
(a)~\includegraphics[keepaspectratio, height=7.5 cm]{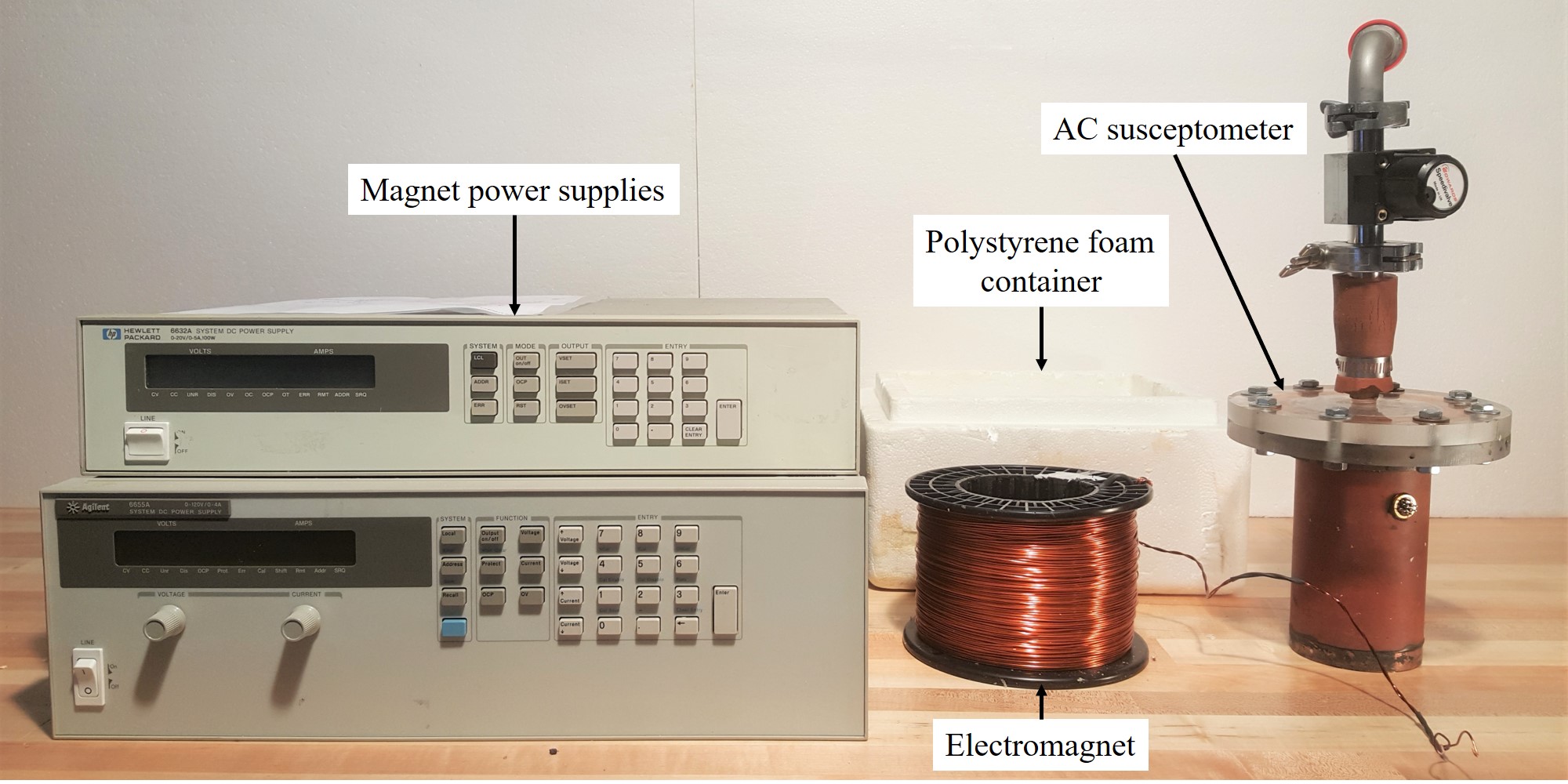}\\
(b)~\includegraphics[keepaspectratio, height=6 cm]{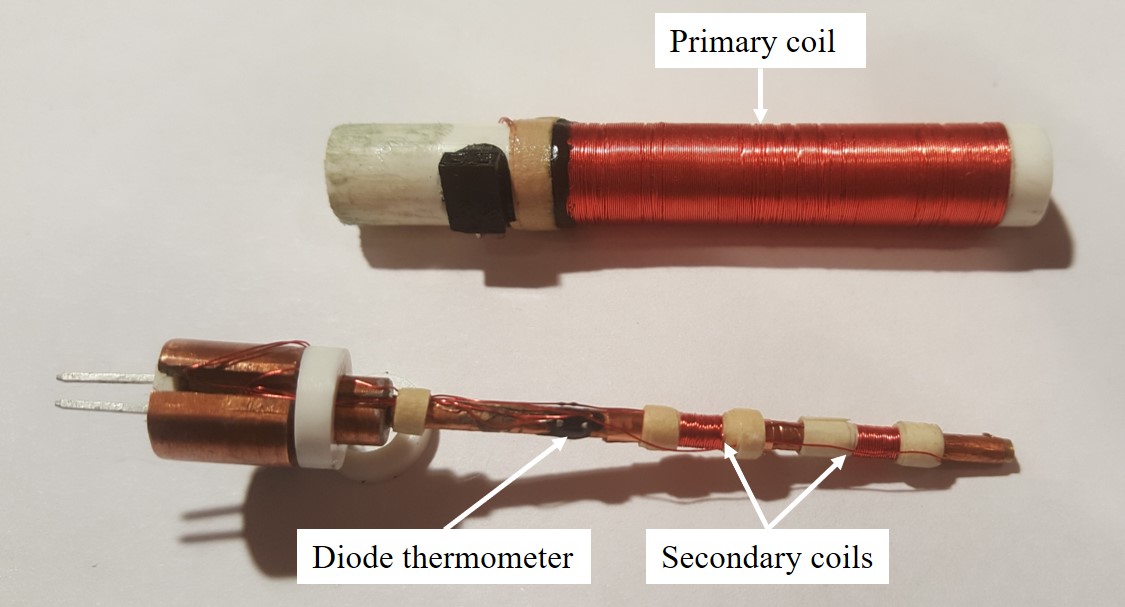}
}
\caption{\label{fig:photo}(a) Photograph of the apparatus.  Not shown are the signal generator used to drive the primary coil of the susceptometer; the amplifier, filter, phase shifter, and lock-in amplifier used to detect the signal from the secondary coils; and the current source and voltmeter used with the diode thermometer. (b) Detailed view of the primary and secondary coils of the susceptmeter.  For scale, the outer diameter of the plastic tube onto which the primary coil was wound is \SI{9.50}{\milli\meter}.  The male connector pins visible in the photograph are connected to the secondary coils.  Not visible in the photograph is a second connector used for the diode thermometer.}
\end{figure}

A schematic diagram of the apparatus is shown in figure~\ref{fig:apparatus}.  The actual apparatus, including a detailed view of the susceptometer coil set, is pictured in figure~\ref{fig:photo}.

\subsection{Susceptometer coil set}
The susceptometer primary and secondary coils were hand wound using \SI{34}{AWG} insulated copper wire.  The primary coil is \SI{40}{\milli\meter} long and was wrapped around a narrow groove that was machined into a plastic tube with inner and outer diameters of \num{5.60} and \SI{9.50}{\milli\meter}, respectively.  The primary has three layers of windings.  The coil leads are soldered to a miniature connector that was attached to the plastic tube using Stycast 2850FT epoxy.   

The pair of counter-wound secondary coils were wrapped onto a \SI{2.05}{\milli\meter} diameter brass rod.  Before winding the coils, the rod was wrapped with copper foil tape to ensure that entire length of the rod maintained a uniform temperature.  Each of the secondary coils has 20 turns.  The superconducting sample is located in the bottom coil closest to the diode used to monitor the sample temperature.  First, a fine file was used to make a recess in the brass rod.  The sample was then held in place using a small amount of Dow Corning silicone vacuum grease. After the sample was in place, the copper tape was wrapped around the recess in the brass rod and the wire was wound such that the sample was centred in the resulting coil.  The samples used in our experiments were typcially $2\times 1.5\times 1~\si{\milli\meter\tothe{3}}$ in size and irregular in shape.  By wrapping one of the secondary coils directly around the sample rod, we achieve a high filling factor and a correspondingly high sensitivity.  Of course, the drawback is that the bottom secondary must be rewound every time the sample is changed.  One could alternatively wind the secondary coils onto a nylon or polyimide tube that is a close fit to the sample rod.  Such a design would maintain a relatively high filling factor while also allowing the coils to be temporarily removed when exchanging samples.

To monitor the sample temperature, an inexpensive surface-mount diode (IN914) was used as a thermometer.  The diode was attached to the to the copper foil-wrapped brass rod using Stycast 2850FT.  As shown in figure~\ref{fig:photo}(b), the diode was positioned close to the bottom secondary coil that contains the sample under investigation.   During measurements, the diode was supplied with a constant current of \SI{100}{\micro\ampere} and the forward-biased voltage drop across the diode was measured.  The diode thermometer was first calibrated against a platinum wire resistance thermometer purchased from Cryogenic Contol Systems, Inc.  The brass rod is soldered into a copper block that presses into the thermal stage used to set the sample temperature.  At the same time the block is pressed into the thermal stage, miniature connectors for the secondary coils and the diode thermometer mate with corresponding connectors permanently attached to the thermal stage.    

\subsection{Thermal stage}\label{sec:thermal}
As shown in figure~\ref{fig:apparatus}, the susceptometer is mounted on a thermal stage that consists of a copper block that has been thermally isolated from the liquid nitrogen bath using a brass rod.  Assuming an isothermal copper block and a brass rod with negligible heat capacity, the relevant thermal time constant is determined from the product of the heat capacity $C_\mathrm{T}$ of the copper and the thermal resistance $R_\mathrm{T}$ of the brass rod.  The thermal resistance is given by:
\begin{equation}
R_\mathrm{T}^{-1}=\frac{A}{\ell}\frac{1}{T_\mathrm{s}-T_\mathrm{b}}\int_{T_\mathrm{b}}^{T_\mathrm{s}}\kappa\left(T\right)\, dT,
\end{equation}
where $\kappa\left(T\right)$ is the temperature-dependent thermal conductivity, $T_\mathrm{s}$ is the sample temperature, $T_\mathrm{b}\approx\SI{77}{\kelvin}$ is the base temperature set by the nitrogen bath, and $A$ and $\ell$ are the cross-sectional area and length of the brass rod, respectively.

The copper block of the thermal stage is a \SI{2.5}{\centi\meter} diameter copper cylinder that is \SI{3.3}{\centi\meter} tall.  The brass rod is a length of \#6--32 threaded ready rod and has a nominal diameter of \SI{3.3}{\milli\meter}.  The length of brass rod between the copper base plate and the copper cylinder at temperature $T_\mathrm{s}$ is $\ell\approx\SI{6}{\milli\meter}$.  Because both the thermal conductivity of brass and specific heat of copper are temperature dependent, the thermal time constant $\tau=R_\mathrm{T}C_\mathrm{T}$ changes as the sample cools from room temperature to \SI{77}{\kelvin}.  Table~\ref{tab:tau} shows the calculated values of $\tau$ at a number of different temperatures.  The thermal conductivity of brass was taken from Ref.~\cite{Powell:1957} and the specific heat of copper was taken from Ref.~\cite{Corruccini:1960}.  
\begin{table}
\caption{\label{tab:tau}Table of thermal time constants as a function of temperature.  $\Delta T\equiv T_\mathrm{s}-T_\mathrm{b}$ and $c_P$ is the volumetric heat capacity of copper.  A base temperature of $T_\mathrm{b}=\SI{77}{\kelvin}$ was assumed.}
\begin{indented}
\item[]\begin{tabular}{cccccc}
\br
$T_\mathrm{s}$ & $c_P$ & $C_\mathrm{T}$ & $\Delta T^{-1}\int\kappa(T)\,dT$ & $R_\mathrm{T}$ & $\tau$\\
(\si{\kelvin}) & (\si{\joule\cm\tothe{-3}\kelvin\tothe{-1}}) & (\si{\joule/\kelvin}) & (\si{\watt\centi\meter\tothe{-1}\kelvin\tothe{-1}}) & (\si{\kelvin/\watt}) & (\si{\second})\\
\hline\hline
\num{300} & \num{3.46} & \num{57.8} & \num{0.698} & \num{10.0} & \num{581}\\
\num{200} & \num{3.19} & \num{53.3} & \num{0.144} & \num{12.0} & \num{640}\\
\num{140} & \num{2.80} & \num{46.9} & \num{0.122} & \num{14.1} & \num{662}\\
\num{100} & \num{2.28} & \num{38.1} & \num{0.430} & \num{16.3} & \num{621}\\
\num{80} & \num{1.84} & \num{30.7} & \num{0.097} & \num{17.8} & \num{546}\\
\br
\end{tabular}
\end{indented}
\end{table}

\begin{figure}[t]
\centering{
\includegraphics[keepaspectratio, width=0.6\columnwidth]{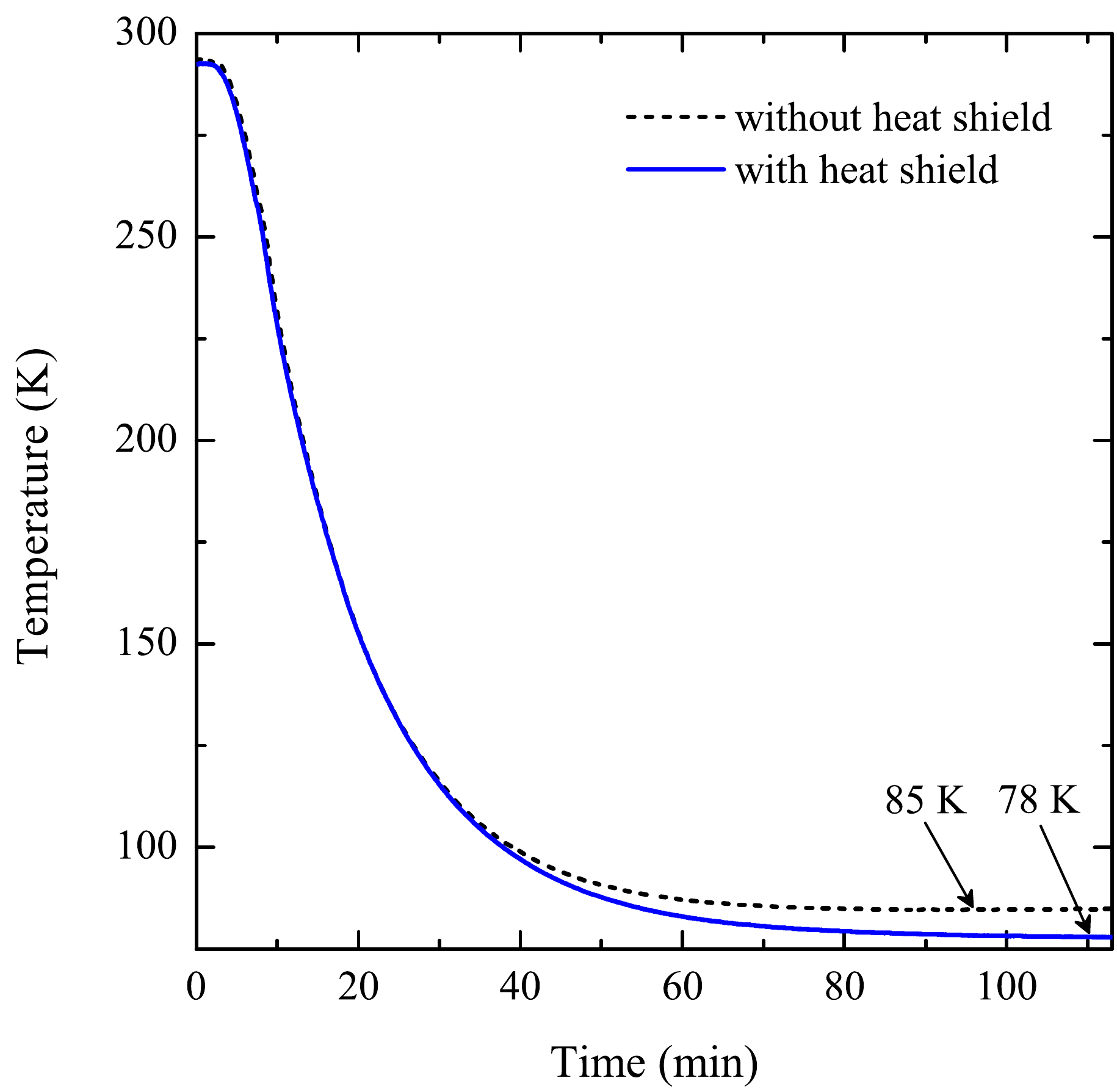}}
\caption{\label{fig:cooling}The sample temperature as a function of time from room temperature to base temperature with (solid line) and without (dashed line) the radiation shield in place.}
\end{figure}

As shown in the table, starting from a temperature of \SI{300}{\kelvin}, the time constant initially increases as the temperature $T_\mathrm{s}$ of the sample stage decreases.  This occurs because the thermal conductivity of brass decreases faster than the specific heat of copper at these temperatures.  However, below about \SI{140}{\kelvin}, copper's specific heat  decreases rapidly causing the thermal time constant to peak and then decrease as $T_\mathrm{s}$ approaches the base temperature.  Over the entire temperature range (\num{77} to \SI{300}{\kelvin}), the calculated time constant changes by a maximum of 20\%.  Starting from room temperature and with $T_\mathrm{b}=\SI{77}{\kelvin}$, figure~\ref{fig:cooling} shows the measured sample temperature as a function of time.
For a thermal $RC$-circuit, the sample is expected to have cooled 95\% of the way to base temperature after a period of three time constants.  From the measured cooling curve, the experimental time constant is found to be approximately \SI{17}{minutes} which is significantly larger than the calculated value that is closer to \SI{10}{minutes}.  The difference is most likely due to the primary coil that slides over top of the secondary coils and thermometer.  The primary coil, wound onto a thermally-insulating plastic former, cools mainly via conduction through the electrical leads.  Therefore, it likely acts as an additional heat load on the sample stage with its own thermal time constant.  Furthermore, the calculated time constant did not take into account of the heat capacities of the AC susceptometer components pictured in figure~\ref{fig:photo}(b).

\subsection{Vacuum chamber}
The bottom of the vacuum chamber is a solid copper plate that is submerged in liquid nitrogen during the measurements.  Using Stycast 2850FT, a paper Bakelite tube was epoxied to a step that was machined around the circumference of the copper plate.  Paper Bakelite was chosen for the walls of the vacuum chamber for its low thermal conductivity allowing it to support a large temperature gradient along its length.  It also has a thermal expansion coefficient that is reasonably well matched to that of copper which prevents the tube from cracking when submerged in liquid nitrogen.  The tube has an outer diameter of \SI{7.6}{\centi\meter}, a wall thickness of \SI{1.3}{\milli\meter}, and a length of \SI{12}{\centi\meter}.  The top of the tube is epoxied to an acrylic ring which mates with an acrylic plate.  The vacuum seal between the acrylic surfaces is made using an o-ring cut from a \SI{0.13}{\milli\meter} thick Kapton sheet.  A thin layer of Dow Corning vacuum grease is used between the Kapton and acrylic surfaces.  A mechanical pump is used to evacuate the vacuum chamber.  A valve attached to the acrylic plate allows the pump to be disconnected after the chamber has been evacuated. 

Near the top of the paper Bakelite tube, a hole was drilled to accept a hermetically-sealed electrical feedthrough.  The nine-pin feedthrough, which accommodates the wiring for the primary and secondary coils and the diode thermometer, was installed using Stycast 2850FT epoxy.  All wiring inside the vacuum chamber was done using a pair of four-wire ribbons of \SI{36}{AWG} phosphor-bronze wire purchased from  Cryogenic Contol Systems, Inc.  Before making connections to components on the sample stage, \SI{20}{\centi\meter} lengths of each ribbon were wrapped around copper heat-sinking posts attached to the copper base plate.  The ribbons were thermally anchored to the heat-sinking posts using GE varnish.

Finally, a polished copper can securely bolted to the \SI{77}{\kelvin} base plate was used as a radiation shield.  The shield completely surrounds the thermal stage supporting the sample and susceptometer.  Its purpose is to suppress radiative heat transfer from the warm walls of the vacuum chamber to the sample stage.  This heat transfer is proportional to $T_\mathrm{vc}^4-T_\mathrm{s}^4$ where $T_\mathrm{vc}$ is the temperature of the vacuum chamber wall.  Figure~\ref{fig:cooling} shows that using the radiation shield lowers the sample base temperature by nearly 10\%.

\subsection{Electromagnet}\label{sec:electromagnet}
Our electromagnet was made from \SI{600}{\meter} of \SI{18}{AWG} wire.  We found that, after cutting out the spindle, our vacuum chamber fit within the inner diameter of the as-purchased spool of wire.  At room temperature, the resistance of the electromagnet was \SI{12.5}{\ohm} and at \SI{77}{\kelvin} it dropped to \SI{1.7}{\ohm}.  To energize the magnet, a parallel combination of an Agilent 6655A (\SI{120}{\volt}/\SI{4}{\ampere}) power supply and a Hewlett-Packard 6632A (\SI{20}{\volt}/\SI{5}{\ampere}) power supply was used.  When the magnet was cold, this combination allowed us to supply \SI{8}{\ampere} of current with \SI{13.6}{\volt}.    

To calibrate the electromagnet, a Texas Instruments DRV5055A2 Hall effect sensor was used.  The sensor was itself first calibrated against an F.\ W.\ Bell model 5070 Teslameter.  At the sample site, a \SI{8}{\ampere} current resulted in a \SI{63}{\milli\tesla} static field along the axis of the susceptometer coils.  The field strength perpendicular to this axis was over 200 times weaker.  The magnet power supplies were controlled via a simple LabVIEW program which allowed us to automate the static magnetic field sweeps when collecting data.  

\subsection{Detection electronics}
\begin{figure}[t]
\centering{
\includegraphics[keepaspectratio, width=0.65\columnwidth]{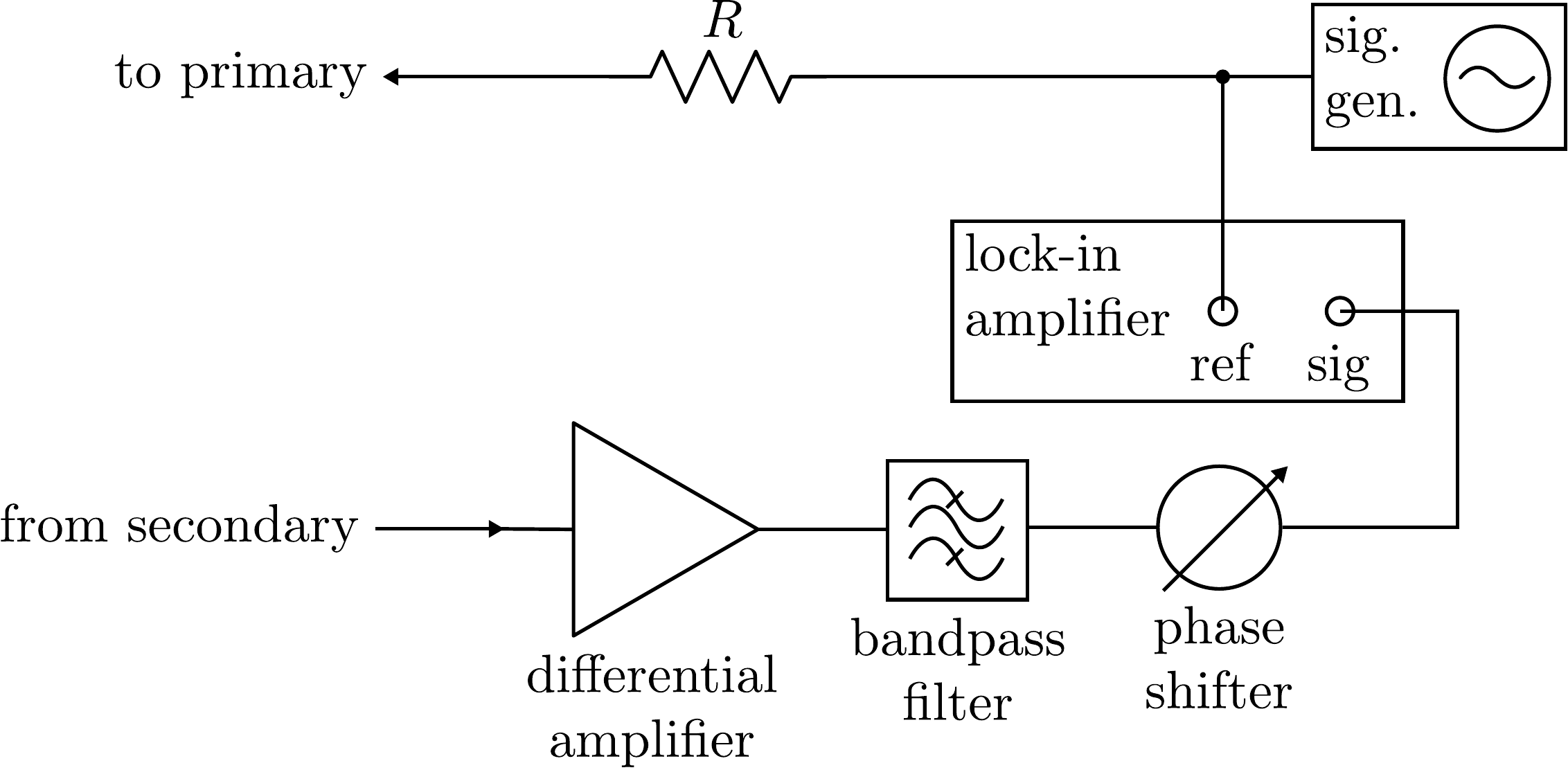}}
\caption{\label{fig:electronics}Schematic diagram of the signal detection electronics.  The primary coil was driven by a signal generator in series with resistance $R$.  The signal from the secondary coils was amplified, filtered, and phase shifted before being detected by a dual-phase lock-in amplifier.}
\end{figure}

To drive the primary coil of the AC susceptometer, a function generator in series with a resistance $R$ was used.  The value of $R$ was chosen to be much greater than the resistance $R_\mathrm{p}$ of the primary coil.  This choice ensures that the primary coil current remains approximately constant as both the coil temperature and $R_\mathrm{p}$ decrease.  As shown in figure~\ref{fig:electronics}, the signal generator output was also used as the reference signal for a Stanford Research Systems SR530 dual-phase lock-in amplifier.

The signal from the counter-wound secondary coils was amplified using a differential amplifier and then passed through a narrow band-pass filter.  Finally, the signal from the filter was phase shifted before being passed to the lock-in amplifier.  The amplifier, filter, and phase shifter were all built into a TeachSpin signal processing unit.  The output of the signal generator was set to \SI{3}{\kilo\hertz} which coincides with the top end of the signal processor's bandwidth.   

Both Refs.~\cite{Yonezawa:2015} and \cite{Nikolo:1995} describe strategies for setting the phase of the signal from the secondary.  The method that we adopted was to adjust the phase to produce a symmetric $\chi^{\prime\prime}$ loss peak through the superconducting transition.  

Finally, we note that the in-phase and out-of-phase signals of the lock-in amplifier were monitored using a pair of Keysight 34401A multimeters.  A third multimeter was used to monitor the forward bias voltage of the diode thermometer.  The data from the multimeters were logged using the same LabVIEW program that controls the DC current of the electromagnet.

\section{Measurement principle}
When $T>T_\mathrm{c}$, the magnetic field due to the primary coil completely penetrates the superconducting sample such that the same magnetic flux passes through the pair of secondary coils.  Because the secondary coils are counter-wound and in series, the net induced emf is zero and a null signal is detected corresponding to $\chi^\prime+\mathrm{i}\chi^{\prime\prime}=0$.  

For $T < T_\mathrm{c}$, the magnetic flux is expelled from the interior of the sample due to the Meissner effect.  The flux expulsion results in a non-zero net induced emf that is detected by the lock-in amplifier.  Because superconductors exhibit perfect diamagnetism deep in the superconducting state, the measured signal from the susceptometer can be calibrated by setting $\chi^\prime=-1$ when $T\ll T_\mathrm{c}$.  In this state, magnetic fields penetrate a depth $\lambda(T)$ into the samples.  For \YBCOo, the low-temperature penetration depth \mbox{$\lambda(0)\approx  \SI{100}{\nano\meter}$} which is typically many orders of magnitude smaller than the sample dimensions~\cite{Pereg:2004}.  

As temperature approaches $T_\mathrm{c}$ from below, flux quanta begin to penetrate the sample in the form of vortex lines resulting in the onset of hysteresis losses associated with the alternating magnetic field~\cite{Wells:2015}.  These losses initially grow as the density of flux lines in the sample increases with increasing temperature. As temperature reaches and then exceeds $T_\mathrm{c}$, the shielding currents and hysteresis losses recede and then vanish~\cite{Nikolo:1995, Giapintzakis:1994}.  The losses are reflected as a peak in the temperature dependence of $\chi^{\prime\prime}$ near $T_\mathrm{c}$.  It is worth noting that the Bean critical-state model can be used to relate the $\chi^{\prime\prime}$ peak to the critical current density of the sample~\cite{Gomory:1997, Bean:1962, Bean:1964}.



\section{Experimental Results}
\subsection{Samples}
\begin{figure}[t]
\centering{
(a)~\includegraphics[keepaspectratio, height=6 cm]{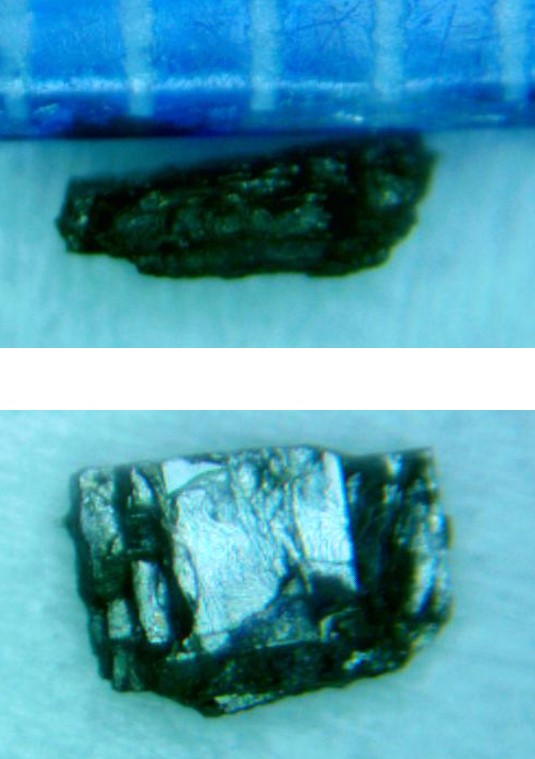}\qquad
(b)~\includegraphics[keepaspectratio, height=6 cm]{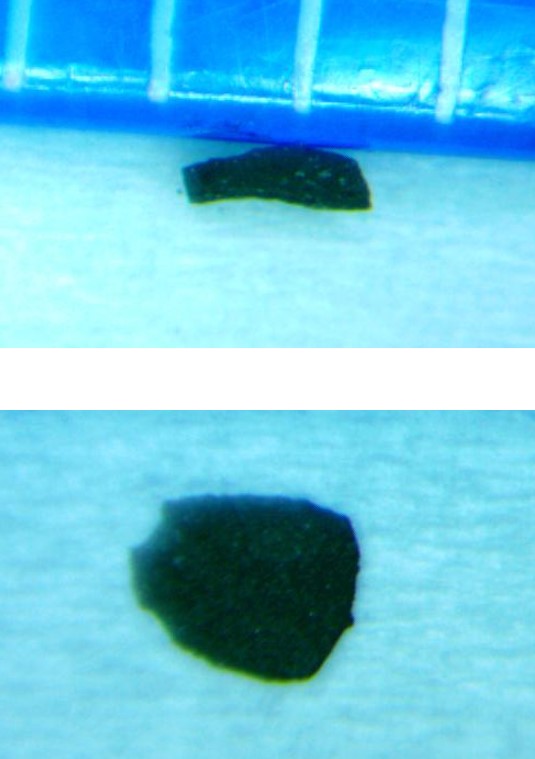}
}
\caption{\label{fig:samples}Photographs of the samples.  The ruler divisions seen in the top panels are \SI{1}{\milli\meter}. (a) The single-crystal sample of \YBCOo{} ($T_\mathrm{c}=\SI{93.1}{\kelvin}$). (b) The sintered polycrystalline \YBCO{} sample ($T_\mathrm{c}=\SI{90.0}{\kelvin})$.}
\end{figure}

We investigated both single-crystal and sintered polycrystalline samples of \YBCO.  The high-quality single-crystal sample was grown by UBC Superconductivity group in yttria-stabilized zirconium oxide (YSZ) crucibles by the \ce{BaO}-\ce{CuO} self-flux method~\cite{Liang:1992}.  After growth, the crystal was annealed in a tube furnace under the flow of high-purity dry oxygen to set to its oxygen content to $y=0.95$ which corresponds to optimal doping and $T_\mathrm{c}\approx\SI{93}{\kelvin}$~\cite{Liang:1998, Liang:2006}.  The polycrystalline sample was purchased from Colorado Superconductor Inc.\ as part of a kit used to demonstrate the levitation of a rare-earth magnet above a superconducting disk.  This sample had a slightly lower superconducting transition temperature of $\approx\SI{90}{\kelvin}$.  Photographs of the two samples used in our measurements are shown in figure~\ref{fig:samples}.

\subsection{Temperature dependence of $\chi^\prime$ and $\chi^{\prime\prime}$}
\begin{figure}[t]
\centering{
\includegraphics[keepaspectratio, width = 0.8\columnwidth]{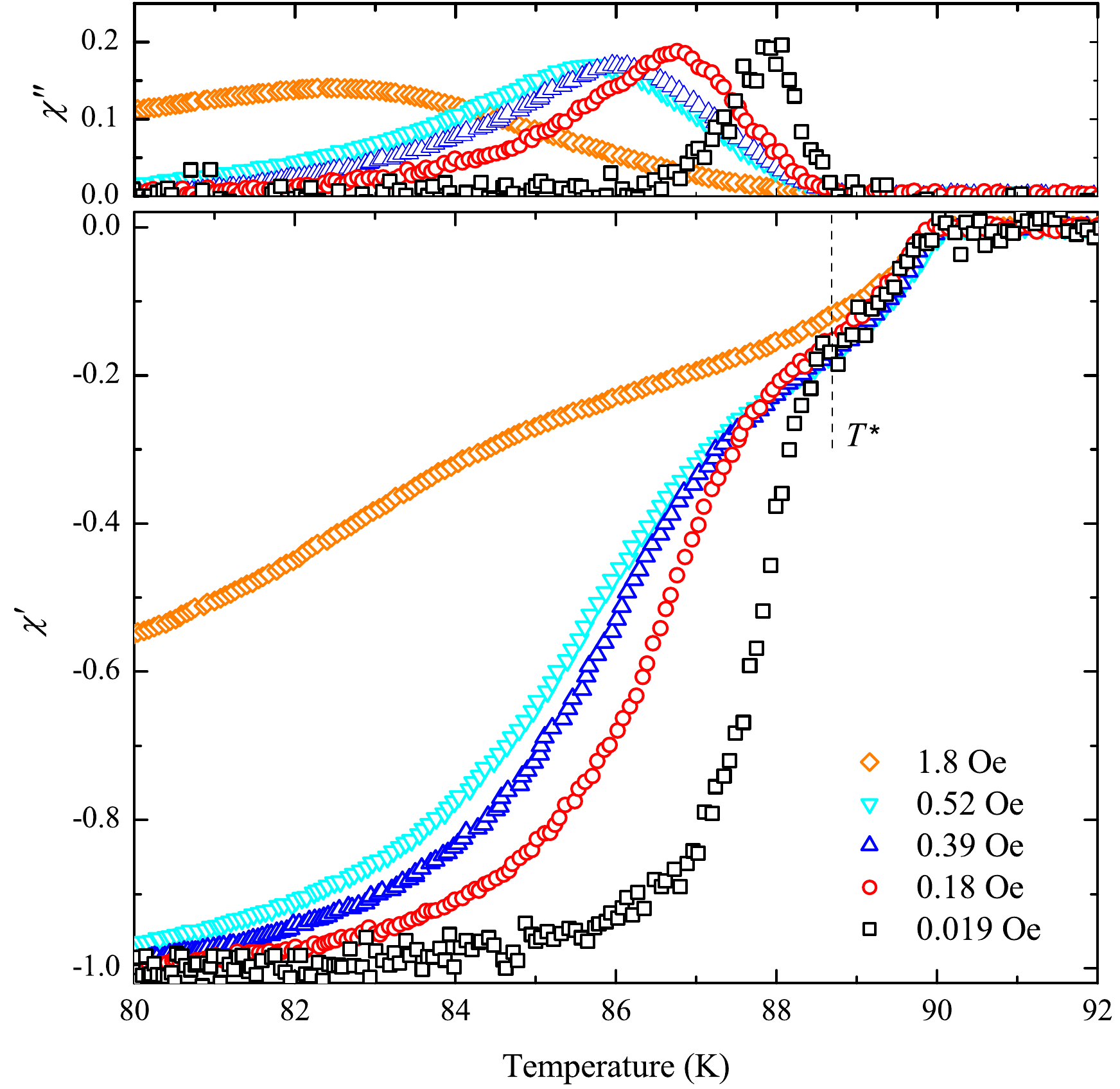}
}
\caption{\label{fig:ChivsTpoly}Temperature dependence of $\chi^\prime$ and $\chi^{\prime\prime}$ for polycrstalline \YBCO{} at five different strengths of the AC field.  The temperature $T^*$ indicated by the dashed line marks an inflection point in the temperature dependence of $\chi^\prime$.}
\end{figure}

\begin{figure}[t]
\centering{
\includegraphics[keepaspectratio, width = 0.8\columnwidth]{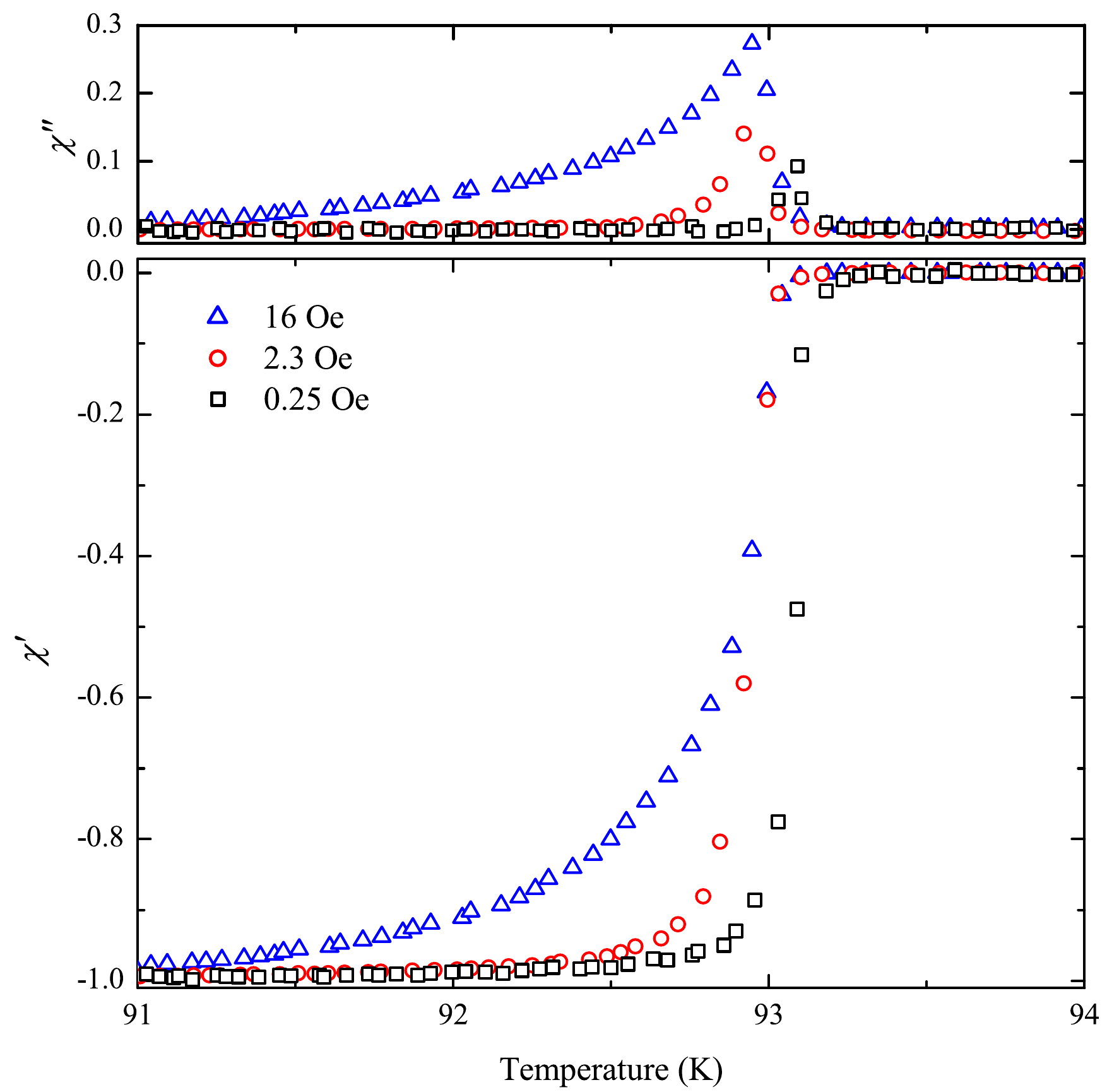}
}
\caption{\label{fig:ChivsTsingle}Temperature dependence of $\chi^\prime$ and $\chi^{\prime\prime}$ for single crystal \YBCOo{} at three different strengths of the AC field.}
\end{figure}

Figures~\ref{fig:ChivsTpoly} and \ref{fig:ChivsTsingle} show the measured temperature dependencies of $\chi^\prime$ and $\chi^{\prime\prime}$ for the polycrystalline and single crystal samples, respectively.
The field strength at the sample site was calibrated using the same Texas Instruments DRV5055A2 Hall effect sensor described in section~\ref{sec:electromagnet}.  As shown in figure~\ref{fig:electronics}, the field strength is set by both the amplitude of the signal generator output and the value of the series resistance $R$.   

The polycrystalline measurements of figure \ref{fig:ChivsTpoly} are qualitatively consistent with the results reported by Nikolo and others using more sophisticated apparatuses~\cite{Nikolo:1995, Dhingra:1993, Sarmago:2004}.  The temperature dependence of $\chi^\prime$ exhibits an inflection point at a temperature $T^*$ just below the superconducting transition temperature.  The response for temperatures $T^*<T<T_\mathrm{c}$ is known as the intrinsic component and the response for $T<T^*$ is known as the coupling component.  The intrinsic component is nearly independent of the AC field strength and is associated with the superconducting grains.  The coupling component, on the other hand, depends strongly on the strength of the AC field -- broadening as the field strength is increased.  This feature is associated with intergranular material which, at low temperatures and fields, couples neighbouring grains.  The intergranular material has a much smaller value of the lower critical field $H_\mathrm{c1}$ than the grains such that flux is able to penetrate into this region much more readily.  

In general, the $\chi^{\prime\prime}$ temperature dependence exhibits both an intrinsic loss peak and a coupling loss peak~\cite{Nikolo:1995}.  However, for the low AC field strengths shown in figure \ref{fig:ChivsTpoly}, the intrinsic peak at $T_\mathrm{c}$ is too small to be clearly observed.  The coupling peak, on the other hand, is very prominent and broadens and shifts to lower temperatures as the AC field strength is increased.

Figure \ref{fig:ChivsTsingle} shows the measured temperature dependencies of $\chi^\prime$ and $\chi^{\prime\prime}$ a for single-crystal sample of \YBCOo{} with the AC fields applied parallel to the $ab$-plane.  At the lowest measurement field, the superconducting transition is extremely sharp with $\chi^\prime$ increasing from \num{-0.9} to \num{-0.1} over a temperature range of just \SI{170}{\milli\kelvin}.  For comparison, at a similar field strength, this transition occurs over a temperature range of \SI{5.8}{\kelvin} for the polycrystalline sample.  Furthermore, the absence of an inflection point in the temperature dependence of $\chi^\prime$ is indicative of a single-domain sample.  Notice also that the dependence of $\chi^\prime$ on the AC field strength is weak compared to that of the polycrystalline sample, even when applying a field strength that is nearly an order of magnitude larger than the maximum value used in figure \ref{fig:ChivsTpoly}.  The $\chi^{\prime\prime}$ measurements exhibit a single intrinsic loss peak that grows in size and broadens as the AC field strength is increased.

\subsection{$H_\mathrm{c1}$ of single-crystal \YBCOo}
This section describes a set measurements used to estimate the lower critical field $H_\mathrm{c1}$ of the \YBCOo{} single crystal at \SI{77}{\kelvin}.  The strategy was to look for small changes in $\chi^\prime$ due to a static magnetic field $H$ applied parallel to the $ab$-plane.  As a measure of the fraction of the sample volume penetrated by magnetic vortices, $\Delta\chi^\prime(H)$ is expected to zero when $H<H_\mathrm{c1}$ and to increase monotonically with $H$ above $H_\mathrm{c1}$.  At \SI{2.3}{Oe} (\SI{0.23}{\milli\tesla}), the AC field strength was chosen to be much smaller than $H_\mathrm{c1}$ while simultaneously maintaining a signal-to-noise ratio sufficient to measure changes in $\chi^\prime$ at the level of a few hundredths of a percent. 

Due to demagnetization effects, magnetic flux can penetrate into the sample at sharp edges and corners at fields less than $H_\mathrm{c1}$~\cite{Liang:1994, Liang:2005, Prozorov:2018}.  Therefore, the measurements described here will not yield an accurate measurement of the intrinsic lower critical field; however, we will be able to clearly identify the field at which our sample enters into the mixed state.  To minimize demagnetization effects, the sample was oriented such that the applied magnetic field was parallel to its long edge.  Furthermore, as shown in figure~\ref{fig:samples}(a), the \YBCOo{} single crystal used in our measurements was selected for its as-grown rounded corners.  

Liang {\it et al.}\ have made accurate measurements of $H_\mathrm{c1}$ on optimally-doped and underdoped \YBCO{} single crystals.  They polished the samples into ellipsoids which ensures that the samples are uniformly magnetized when immersed in a uniform magnetic field~\cite{Prozorov:2018}.  Furthermore, the polishing introduces a surface roughness that suppresses the so-called Bean-Livingston (BL) surface barrier which, for samples with smooth surfaces, can delay the initial flux penetration until fields above $H_\mathrm{c1}$ are reached~\cite{Liang:1994, Liang:2005, BL:1964}. 

\begin{figure}[t]
\centering{
\includegraphics[keepaspectratio, width = 0.8\columnwidth]{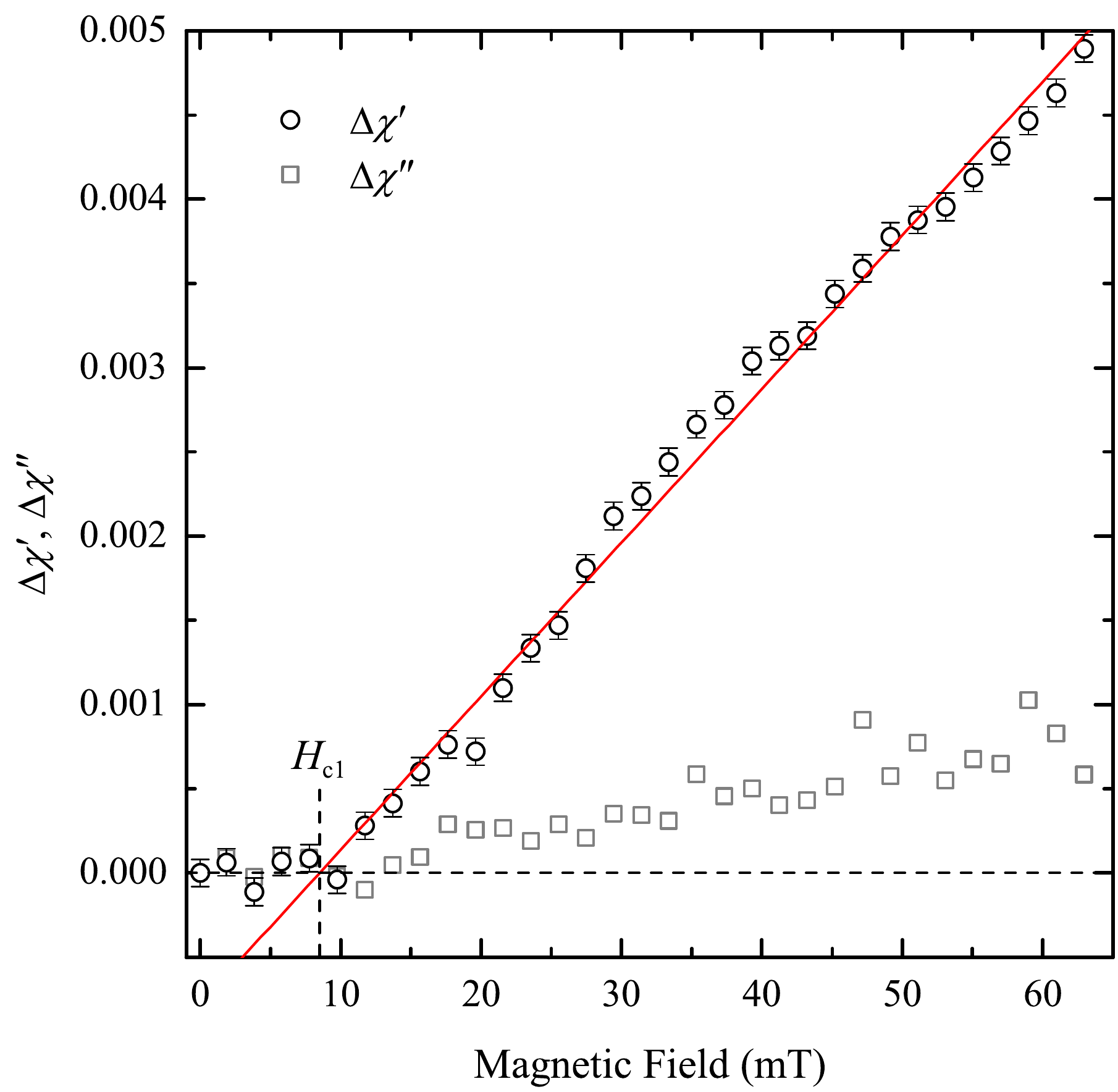}
}
\caption{\label{fig:Hc1}Change in $\chi^\prime$ (circles) and $\chi^{\prime\prime}$ (squares) of the \YBCOo{} single crystal as a function of the applied static magnetic field strength.  The sample temperature was \SI{77}{\kelvin} and the field was applied parallel to the sample's $ab$-plane.  The solid line is a fit to the $\Delta\chi^\prime$ data for $H>\SI{10}{\milli\tesla}$.}
\end{figure}

Figure~\ref{fig:Hc1} shows the measured field dependence of the real and imaginary components of $\Delta\chi$ for the \YBCOo{} single crystal at \SI{77}{\kelvin}.
These data were collected by first allowing the sample to cool to a stable base temperature.  We then measured average values of $\chi$ in an applied field $H$ and in zero field.  Zero-field averages were taken both before $\left(\chi_1(0)\right)$ and after $\left(\chi_2(0)\right)$ measurements at a nonzero field $\left(\chi(H)\right)$.  The change in susceptibility was then calculated using:
\begin{equation}
\Delta\chi(H)=\chi(H)-\frac{1}{2}\left[\chi_1(0)+\chi_2(0)\right].
\end{equation}   
For static fields below $\approx\SI{10}{\milli\tesla}$, $\Delta\chi^\prime$ was found to be zero.  It then increased linearly with the field strength such that 0.5\% of the sample was penetrated by magnetic vortices when $H=\SI{63}{\milli\tesla}$.  Although small in magnitude, figure~\ref{fig:Hc1} also shows a linear change in $\chi^{\prime\prime}$ as function of $H$.  It should be noted that the observed field dependence of $\Delta\chi^{\prime\prime}$ could, in part, be due to an improperly set phase angle which would contaminate the channel of the lock-in amplifier used to monitor $\chi^{\prime\prime}$ with a small fraction of the $\chi^\prime$ signal.  After the measurements at base temperature, we verified that there was no systematic change in $\chi^\prime$ and $\chi^{\prime\prime}$ as a function of field strength when the \YBCOo{} sample was in its normal state ($T>\SI{100}{\kelvin}$).   The solid line in figure~\ref{fig:Hc1} is a fit to the $\Delta\chi^\prime$ data above \SI{10}{\milli\tesla}.  $H_\mathrm{c1}$ was estimated to be \SI{8.50}{\milli\tesla} by finding the field at which the fit line intersects $\Delta\chi^\prime=0$.  

Our estimate of $H_\mathrm{c1}$ is significantly larger than the \SI{77}{\kelvin} value of \SI{4.7}{\milli\tesla} that Liang {\it et al.}\ found for \YBCOo{} when working with a polished ellipsoid sample.  One possible reason for the observed difference is the aforementioned BL surface barrier.  The as-grown crystal that we measured, shown in figure~\ref{fig:samples}(a), had no surface preparation to suppress this effect.  Sample misalignment is another possible reason for the larger-than-expected $H_\mathrm{c1}$ measurement.  At \SI{77}{\kelvin}, the lower critical field of \YBCOo{} with $H$ parallel to the $c$-axis is nearly five times larger than the corresponding value with $H$ parallel to the $ab$-plane~\cite{Liang:1994}.  Therefore, even a small misalignment between the applied magnetic field and the $ab$-plane of the sample would lead to an enhancement of the apparent $H_\mathrm{c1}$.   

Because our apparatus did not allow the sample temperature to be regulated above the system base temperature, we were unable to reliably measure the temperature dependence of $H_\mathrm{c1}$ between \SI{77}{\kelvin} and $T_\mathrm{c}$.  However, by sweeping the static field strength quickly while allowing the sample temperature to drift from $T_\mathrm{c}$ towards \SI{77}{\kelvin}, we were able to measure the temperature dependence of the slope of the $\Delta\chi^\prime$ versus $H$ data.  These results are presented in the supplementary material (available online).  We found that, at a sample temperature of \SI{84}{\kelvin}, $d\chi^\prime/dH$ was more than double its value of \SI{0.091}{\tesla\tothe{-1}} at \SI{77}{\kelvin}.  This analysis is useful because, despite being unable to determine the temperature dependence of $H_\mathrm{c1}$, it allows students to deduce that the superconducting state is more susceptible to penetration by magnetic flux as $T_\mathrm{c}$ is approached.  

\subsection{Mixed state of polycrystalline \YBCO}
\begin{figure}[t]
\centering{
\includegraphics[keepaspectratio, width = 0.8\columnwidth]{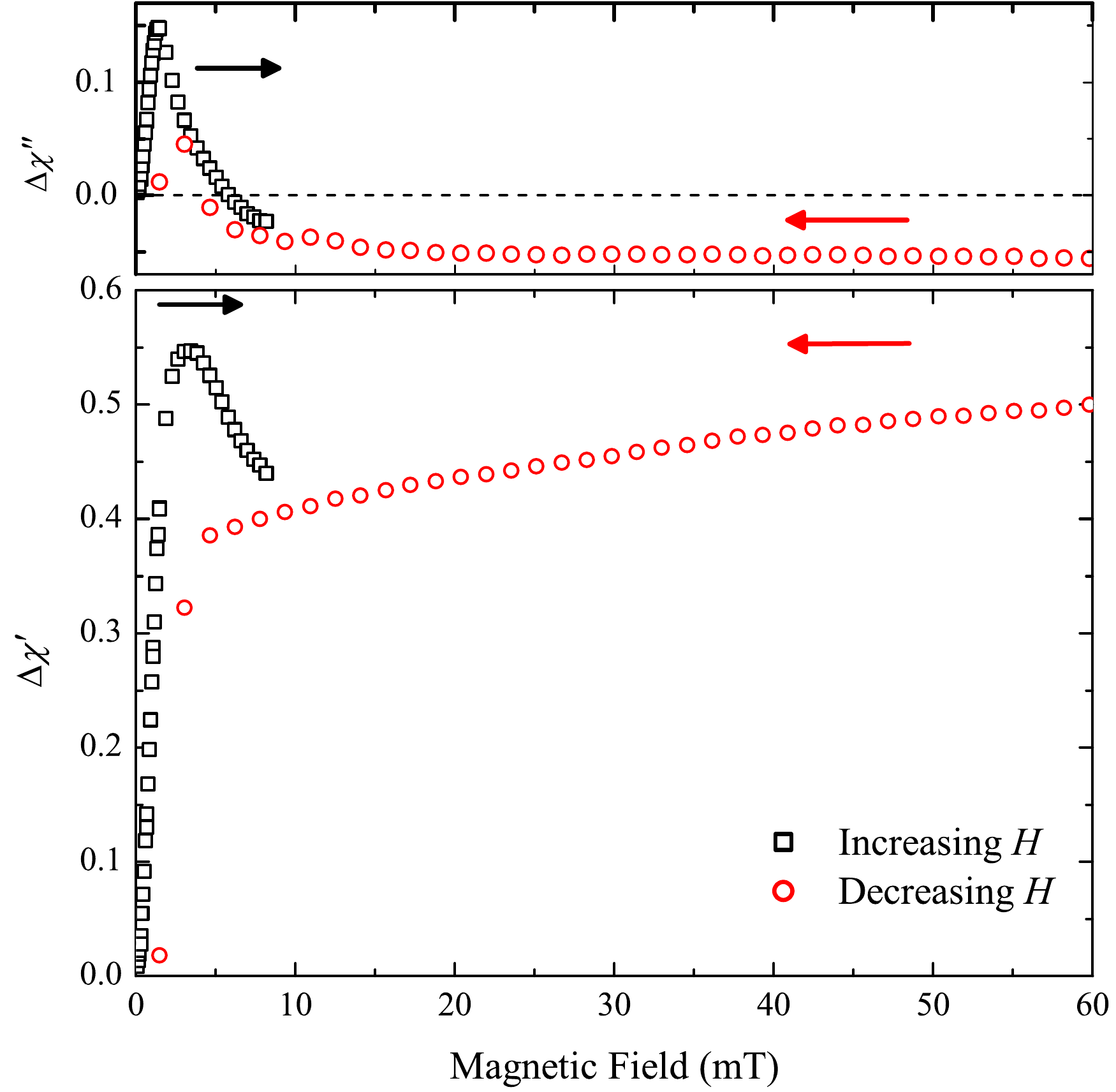}
}
\caption{\label{fig:polyDeltaChi}$\Delta\chi^\prime$ and $\Delta\chi^{\prime\prime}$ of the polycrystalline sample of \YBCO{} as a function of $H$.  The square data points were taken with $H$ increasing and the circular data were taken with $H$ decreasing. $\Delta\chi^\prime$ exhibits two distinct slopes.  The steep slope at low fields is associated with the weakly-superconducting intergranular material and the shallower slope at higher fields is associated with the superconducting grains.  $\Delta\chi^{\prime\prime}$ peaks at low fields and becomes negative at higher field strengths.}
\end{figure}

As shown in figure~\ref{fig:polyDeltaChi}, we also measured the field dependencies of $\Delta\chi^\prime$ and $\Delta\chi^{\prime\prime}$ of the polycrystalline sample of \YBCO.
For these measurements, the AC field strength was set to \SI{0.18}{Oe}.  Data were collected with both an increasing (square data points) and decreasing (circular data points) static field strength.  We observed two distinct slopes in the $\Delta\chi^\prime$ data.  The steep slope at low field strengths ($H<\SI{5}{\milli\tesla}$) is associated with the weakly-superconducting intergranular material.  With $H$ increasing, a peak in $\Delta\chi^\prime$ was observed.  The origin of this peak is not currently understood and it was not observed in the data collected with $H$ decreasing.  At higher fields, the continued increase in $\Delta\chi^\prime$ with increasing $H$ is associated with magnetic flux penetration into the superconducting grains.  Comparing figures~\ref{fig:Hc1} and \ref{fig:polyDeltaChi} reveals that the value of $d\chi^\prime/dH$ for the polycrystalline sample when $H>\SI{10}{\milli\tesla}$ is approximately \num{20} times greater than that of the single-crystal sample.  This observation indicates that even the superconducting grains of the polycrystalline are far more susceptible to penetration by magnetic vortices than the single-crystal sample of \YBCOo.

$\Delta\chi^{\prime\prime}$ exhibits a low-field peak and becomes negative at high fields for both the increasing- and decreasing-$H$ datasets.  These effects are presumably due to non-negligible zero-field losses associated with the weakly-superconducting intergranular material.  As the field strength is increased, $T_\mathrm{c}$ of the intergranular material is diminished and the losses increase (see figure~\ref{fig:ChivsTpoly}).  Eventually, the field strength becomes sufficient to completely suppress the weak intergranular superconductivity, and the associated losses, such that $\Delta\chi^{\prime\prime}<0$.

\section{Summary}
We have described a simple, yet sensitive, AC susceptometer designed for the undergraduate laboratory.  The system base temperature is \SI{77}{\kelvin} and it does not require temperature regulation via a PID controller.  Instead, materials are chosen to set the thermal time constant of the sample stage such that the cooling time is suitable for a three-hour undergraduate laboratory.  We used the susceptometer to study single-crystal and polycrystalline samples of the high-temperature superconductor \YBCO.  By wrapping one of the secondary coils directly onto the sample support rod we achieved a high experimental sensitivity.  The measured superconducting transition of the single-crystal \YBCOo{} sample was very sharp and relatively robust against changes to the strength of AC magnetic field supplied by the primary coil.  On the other hand, the polycrystalline sample exhibited two distinct transitions separated by an inflection point.  The broad transition associated with the intergranular material was very sensitive to the strength of the AC field.  

To go beyond a relatively straightforward characterization of the superconducting transition, the vacuum chamber enclosing our compact susceptometer was inserted into the bore of an electromagnet used to expose the samples to a static magnetic field.  Using this setup, we were able to estimate the lower critical field $H_\mathrm{c1}$ of the single-crystal sample.  By tracking $d\chi^\prime/dH$ as a function of the sample temperature, we were also able to demonstrate that the superconducting state is more susceptible to penetration by magnetic flux as $T_\mathrm{c}$ is approached.

\section*{Acknowledgments}
We gratefully acknowledge the UBC Superconductivity group for supplying the high-quality single-crystal sample of \YBCOo{}.  We also thank R Liang, J C Baglo, P Dosanjh, D A Bonn and W N Hardy of the Superconductivity group for enlightening discussions.

\clearpage
\title[Supplementary Material]{The Superconducting Transition and Mixed State of \YBCOo: Supplementary Material}
\author{Zhongda Huang, Yihang Tong and Jake S Bobowski} 
\address{Department of Physics, University of British Columbia, Kelowna, British Columbia, Canada V1V 1V7} 
\ead{jake.bobowski@ubc.ca}

\begin{abstract}
This supplementary material presents a measurement of the temperature dependence of $d\chi^\prime/dH$ for a single-crystal sample of \YBCOo{} with $T_\mathrm{c}=\SI{93.1}{\kelvin}$ and a static magnetic field $H$ applied parallel to the sample's $ab$-plane.  The measurements were made over a temperature range of \num{77} to \SI{85}{\kelvin} using a simple AC susceptometer built in-house.    
\end{abstract}

\setcounter{equation}{0}
\setcounter{figure}{0}
\setcounter{table}{0}
\setcounter{section}{0}
\setcounter{page}{1}
\makeatletter
\renewcommand{\theequation}{S\arabic{equation}}
\renewcommand{\thefigure}{S\arabic{figure}}
\renewcommand{\thetable}{S\arabic{table}}
\renewcommand{\thesection}{S\Roman{section}}
\renewcommand{\thepage}{S\arabic{page}}  

\begin{figure}[h!]
\centering{
\includegraphics[keepaspectratio, height=9cm]{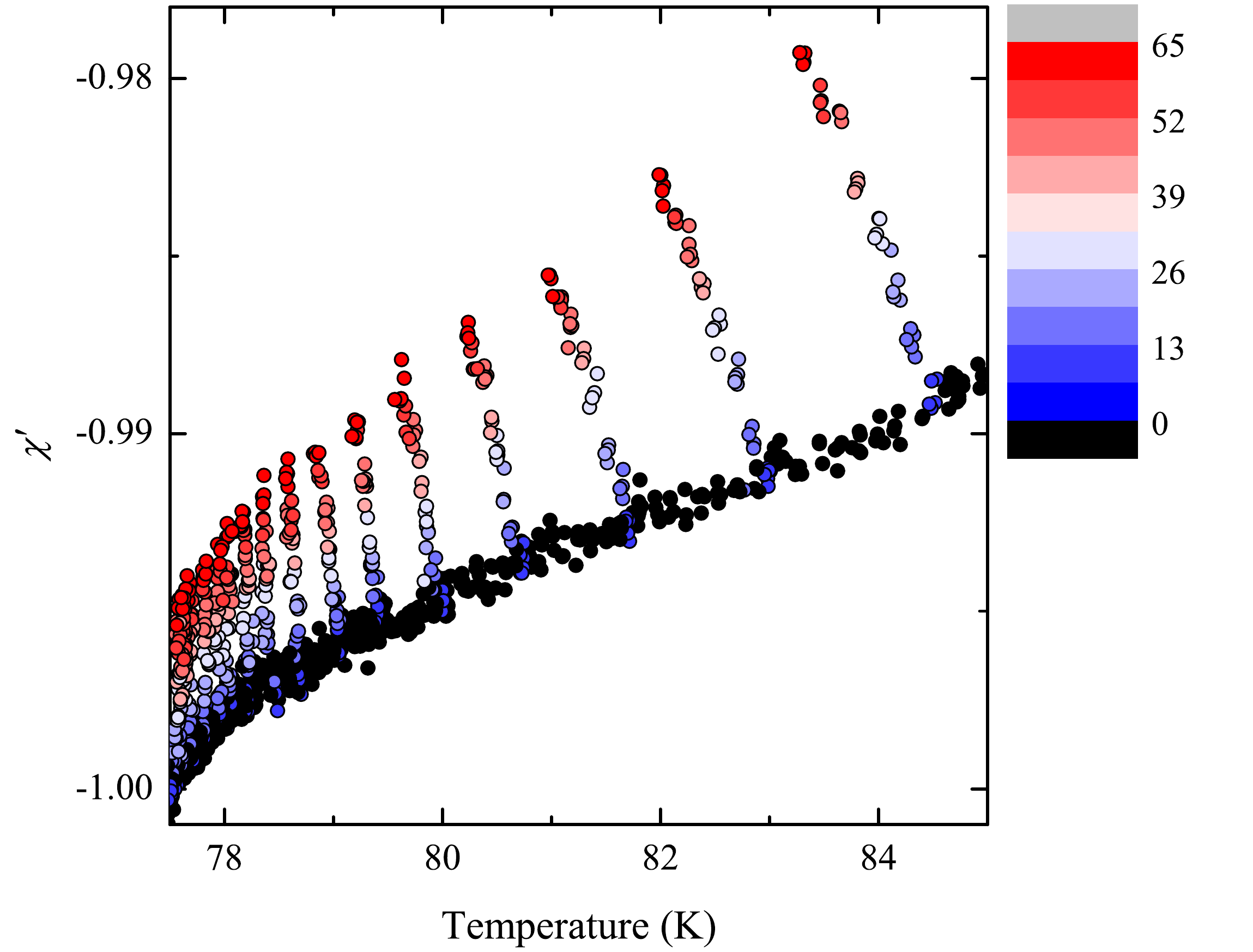}
}
\caption{\label{fig:chi1drift}$\chi^\prime$ as a function of temperature as the strength of the applied static magnetic field was cycled between zero and a set of nonzero values.  The colour scale on the right gives the value of $H$ in milli-Tesla.}
\end{figure}

For these measurements, the temperature of the \YBCOo{} sample was initially just below $T_\mathrm{c}$ and cooling.  The cooling rate was set by the thermal circuit described in section~\ref{sec:thermal} and table~\ref{tab:tau} of the main manuscript.  We then repeatedly cycled the static magnetic field between zero and a set of nonzero values as the sample drifted towards its base temperature.  This sequence of measurements was automated using a simple LabVIEW program which allowed us to track $\Delta\chi^\prime(H)$ as a function of temperature.  The raw data collected is shown in figure~\ref{fig:chi1drift}.   

Because there was no temperature control, it was necessary to sweep through the set of $H$ values quickly to limit the temperature change during each sweep.  For example, the slope of spike in $\chi^\prime$ at \SI{84}{\kelvin} in figure~\ref{fig:chi1drift} shows that the sample temperature changed by more than \SI{1}{\kelvin} while $H$ was cycled through values spanning zero to \SI{63}{\milli\tesla}.  Closer to the base temperature of \SI{77}{\kelvin}, the temperature drift rate is substantially lower such that the spikes representing individual field sweeps appear more vertical.  Because there was limited averaging time, it was not possible to reliably estimate the temperature dependence of $H_\mathrm{c1}$ from these data.  However, it was possible to construct plots of $\Delta\chi^\prime$ versus $H$, similar to figure~\ref{fig:Hc1} of the main manuscript, for each of the field sweeps.  These plots were then used to determine the temperature dependence of $d\chi^\prime/dH$ which is a measure of the fragility of the superconducting state to external magnetic fields. 

\begin{figure}[t]
\centering{
\includegraphics[keepaspectratio, width = 0.6\columnwidth]{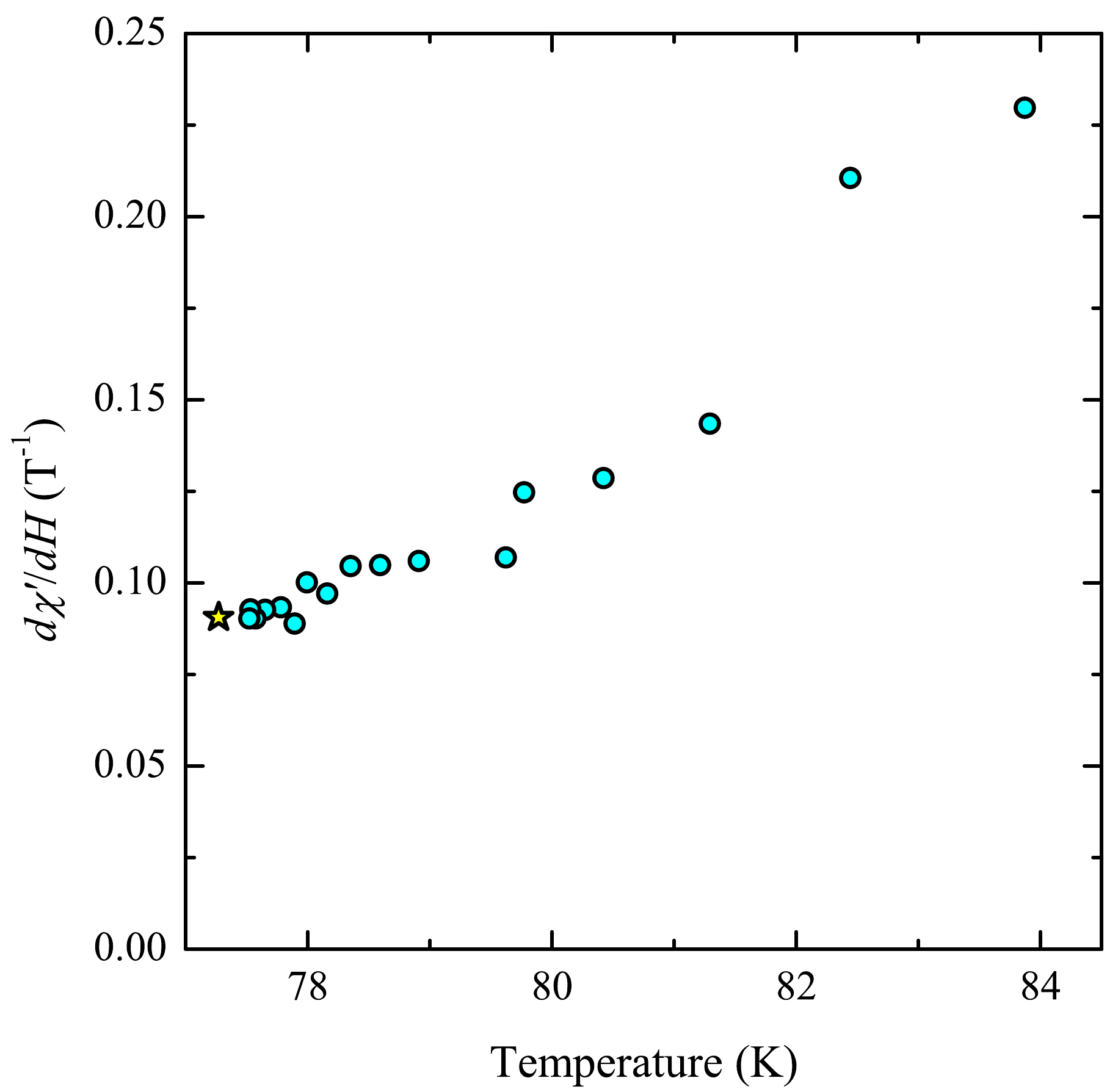}
}
\caption{\label{fig:dXdH}Plot of $d\chi^\prime/dH$ as a function of temperature.  The star data point corresponds to the slope of the solid line shown in figure~\ref{fig:Hc1} of the main manuscript.  The circular points were determined from an analysis of the data shown in figure~\ref{fig:chi1drift}.}
\end{figure}

Figure~\ref{fig:chi1drift} shows that $\Delta\chi^\prime$ at the highest field strengths tested increases as temperature approaches $T_\mathrm{c}$.  This same observation is reflected in the plot of $d\chi^\prime/dH$ versus temperature shown in figure~\ref{fig:dXdH}.
This relatively simple analysis of the data confirms that, as expected, the superconducting state is more susceptible to penetration by magnetic flux as $T_\mathrm{c}$ is approached.


\end{document}